\documentclass[onecolumn,10pt]{article}
\usepackage{amsmath}
\usepackage{amsfonts}
\usepackage{amssymb}
\usepackage{amsthm}
\usepackage{times}
\usepackage{cancel}
\usepackage[dvips]{graphics,color}

\newcommand{\trace}[1]{\mbox{$\mathrm{Tr}$}(#1)}
\newcommand{\ptrace}[2]{\mbox{$\mathrm{Tr}_{#1}$}(#2)}

\newcommand{\hil}[1]{\mbox{$\mathcal{#1}$}}

\newcommand{\ket}[1]{| #1 \rangle}
\newcommand{\bra}[1]{\langle #1 |}

\usepackage{tikz}

\title{\textbf{Quantum Probabilities: \\An Information-Theoretic Interpretation\footnote{Forthcoming in Stephan Hartmann and Claus Beisbart (eds.), Probabilities in Physics (Oxford University Press, 2010)}}}
\author{Jeffrey Bub\\ \small \textit{Philosophy Department and Institute for Physical Science and Technology}\\  \small \textit{University of Maryland, College Park, MD 20742, USA}}
\date{}
\normalsize
\begin{document}

\maketitle

\begin{abstract}
This Chapter develops a realist information-theoretic interpretation of the nonclassical features of quantum probabilities.  On this view, what is fundamental in the transition from classical to quantum physics is the recognition that \emph{information in the physical sense has new structural features}, just as the transition from classical to relativistic physics rests on the recognition that space-time is structurally different than we thought. Hilbert space, the event space of quantum systems, is interpreted  as a kinematic (i.e., pre-dynamic) framework for an indeterministic physics, in the sense that the geometric structure of Hilbert space imposes objective probabilistic or information-theoretic constraints on correlations between events, just as  the geometric structure of Minkowski space in special relativity imposes spatio-temporal kinematic constraints on events. The interpretation of quantum probabilities is more subjectivist in spirit than other discussions in this book (e.g., the chapter by Timpson), insofar as the quantum state is interpreted as a credence function---a bookkeeping device for keeping track of probabilities---but it is also objective (or intersubjective), insofar as the credences specified by the quantum state are understood as uniquely determined, via Gleason's theorem, by objective correlational constraints on events in the nonclassical quantum event space defined by the subspace structure of Hilbert space.     
\end{abstract}

\bigskip

\section{Introduction}

Quantum probabilities are puzzling because quantum correlations are puzzling, and quantum correlations are puzzling in the way they differ from classical correlations. The aim of this Chapter is to argue for a realist information-theoretic interpretation of the nonclassical features of quantum probabilities. On this view, the transition from classical to quantum physics rests on the recognition that physical information, in Shannon's sense \cite{Shannon}, is structurally different than we thought, just as the transition from classical to relativistic physics rests on the recognition that space-time is structurally different than we thought.  Hilbert space, the event space of quantum systems, is interpreted  as a kinematic (i.e., pre-dynamic) framework for an indeterministic physics, in the sense that the geometric structure of Hilbert space imposes objective probabilistic or information-theoretic constraints on correlations between events, just as in special relativity the geometric structure of Minkowski space imposes spatio-temporal kinematic constraints on events. 

The difference between classical correlations and nonclassical correlations can be brought out simply in terms of 2-person games.  I discuss such games  in \S 2, and in \S 3 I focus on quantum probabilities. It turns out that the irreversible loss of information in quantum conditionalization---the  `irreducible and uncontrollable disturbance' involved in a quantum measurement process, to use Bohr's terminology---is a generic feature of nonclassical probabilistic theories that satisfy a `no signaling' constraint. 

`No signaling' is the requirement that no information should be available in the marginal probabilities of measurement outcomes in a region $\mathbf{A}$ about alternative choices made by an agent in region $\mathbf{B}$. For example, an observer, Alice, in region $\mathbf{A}$ should not be able to tell what observable Bob measured in region $\mathbf{B}$, or whether Bob performed any measurement at all, by looking at the statistics of her measurement outcomes, and conversely.  Formally, if Alice measures the observable $A$ with outcomes $a$ in some set and Bob measures the observable $B$ with outcomes $b$ in some set, the constraint is: \begin{eqnarray}
\sum_{b}p(a,b|A,B) \equiv p(a|A,B) & = & p(a|A),\,\mbox{for all}\, B \label{eqn:nosignal1} \\
\sum_{a}p(a,b|A,B) \equiv p(b|A,B) & = & p(b|B),\,\mbox{for all}\, A \label{eqn:nosignal2} 
\end{eqnarray}
Here $p(a,b|A,B)$ is the probability of obtaining the pair of outcomes $a, b$ in a joint measurement of the observables $A$ on system $\mathbf{A}$ and $B$ on system $\mathbf{B}$, $p(a|A,B)$ is the marginal probability of obtaining the outcome $a$ for $A$ when $B$ is measured in region $\mathbf{B}$, and $p(b|A,B)$ is the marginal probability of obtaining the outcome $b$ for $B$ when $A$ is measured in region $\mathbf{A}$. The `no signaling' constraint requires the marginal probability  $p(a|A,B)$ to be independent of the choice of measurement performed on system $\mathbf{B}$ (and independent of whether system $\mathbf{B}$ is measured at all), i.e.,  $p(a|A,B) = p(a|A)$, and similarly for the marginal $p(b|A,B)$ with respect to measurements on system $\mathbf{A}$, $p(b|A,B) = p(b|B)$. 

In \S 3, I show that the quantum `measurement disturbance' is an unavoidable consequence of the non-existence of  a universal cloning device for nonclassical extremal states representing multipartite probability distributions. Such a device would allow signaling and so is excluded by the `no signaling' constraint.  

In \S4, following Pitowsky \cite{Pitowsky07},  I distinguish two measurement problems, a `big' measurement problem and a `small' measurement problem. I sketch a solution to the `small' measurement problem as a consistency problem, exploiting the phenomenon of decoherence, and I argue that the `big' measurement problem is a pseudo-problem that arises if we take the quantum pure state as the analogue of the classical pure state, i.e., as a representation of physical reality, in the sense that the quantum pure state is the `truthmaker' for propositions about the occurrence and non-occurrence of events. 

Finally, in \S 5, I clarify the sense in which the  information-theoretic interpretation here is proposed as a realist interpretation of quantum mechanics. The interpretation of quantum probabilities is more subjectivist in spirit than other discussions in this book (e.g., the chapter by Timpson), insofar as the quantum state is interpreted as a credence function---a bookkeeping device for keeping track of probabilities. Nevertheless, the interpretation is objective (or intersubjective), because the credences specified by the quantum state are understood as uniquely determined, via Gleason's theorem \cite{Gleason}, by objective correlational constraints on events in the nonclassical quantum event space defined by the subspace structure of Hilbert space. On this view, in the sense of Lewis's Principal Principle, Gleason's theorem relates an objective feature of the world, the nonclassical structure of objective chances, to the credence function of a rational agent. The notion of objective chance can be understood in the metaphysically `thin' Humean or Lewisian sense  outlined by Hoefer \cite{Hoefer2007}, and Frigg and Hoefer \cite{FriggHoefer2009}, for whom chances are not irreducible modalities, or propensities, or necessary connections, but simply  features of the pattern of actual events: numbers satisfying probability rules that are part of the best system of such rules, in the sense of simplicity, strength, and fit, characterizing the `Humean mosaic,' the collection of everything that actually happens at all times.\footnote{Note that this theory of objective chance differs in important respects from Lewis's own account in \cite{Lewis1980}, while following Lewis in broad outline. For details, see \cite{Hoefer2007}. This is one way of spelling out what the objective probabilities might be---it is not a necessary component of the information-theoretic interpretation.}

The discussion draws on my analysis of quantum probabilities in \cite{Bub2007}, and on joint work with Itamar Pitowsky in \cite{BubPitowsky2010} and Allen Stairs in \cite{BubStairs2009}.

\section{Classical and Nonclassical Correlations}

To bring out the difference between classical and nonclassical correlations, consider the following game between two players, Alice and Bob, and a moderator. The moderator supplies Alice and Bob with a prompt, or an input, at each round of the game, where these inputs are selected randomly from a set of possible inputs, and Alice and Bob are supposed to respond with an output, either 0 or 1, depending on the input. They win the round if their outputs for the given inputs are correlated in a certain way. They win the game if they have a winning strategy that guarantees a win on each round.\footnote{Note that they could win an arbitrary number of rounds purely by chance without any winning strategy.} Alice and Bob are told what the inputs will be (i.e., the set of possible inputs for Alice, and the set of possible inputs for Bob), and what the required correlations are, i.e., what counts as winning a round. They are allowed to confer on a joint strategy before the game starts, but once the game starts they are separated and not allowed further communication

Denote the inputs by $x$ and $y$ for Alice and Bob, respectively, and the outputs by $a$ and $b$, respectively. Suppose the inputs for Alice and for Bob are random and can take two values, 0 or 1. The winning correlations are as follows (where `$\cdot$' denotes the Boolean product and `$\otimes$' denotes the Boolean sum, or addition mod 2):
\begin{itemize}
\item[1.] if the inputs $x, y$ are 00, 01, or 10, then $a\cdot b = 0$ (i.e., the outputs $a, b$ are never both 1)
\item[2.]  if the inputs $x, y$ are 11, then $a \oplus b = 0$ (i.e., the outputs $a, b$ are the same, either both 0 or both 1)
\end{itemize}

Since no communication is allowed between Alice and Bob after the game starts, Alice cannot know Bob's inputs, and so she cannot choose the distribution of her outputs over consecutive rounds of the game to reflect Bob's input, and conversely. It follows that the marginal probabilities satisfy the `no signaling' constraint:
\begin{eqnarray}
\sum_{b\in\{0,1\}}p(a,b|x,y)  =  p(a|x), \, a, x, y \in\{0,1\} \\
\sum_{a\in\{0,1\}}p(a,b|x,y)   =  p(b|y), \, b, x, y \in\{0,1\}
\end{eqnarray}
i.e.,  the marginal probabilities of Alice's outputs will be independent of Bob's choice of input, and conversely.

Alice and Bob are supposed to be symmetrical players, so we require the marginal probabilities to be the same for Alice and for Bob, and we also require that the marginal probability of a particular output for a player should be independent of the the same player's input, i.e., we require:
\begin{itemize}
\item[3.] $p(a=0|x=0) = p(a=0|x=1) = p(b=0|y=0) = p(b=0|y=1)$\\
$p(a=1|x=0) = p(a=1|x=1) = p(b=1|y=0) = p(b=1|y=1)$
\end{itemize}
It follows that the marginal probability of a particular output for a player is independent of the player, and independent of either player's input.

Denote the marginal probability of output 1 by $p$. The winning correlations are summed up in Table 1. 
\begin{table}[h!]
\begin{center}
\begin{tabular}{|ll||ll|ll|} \hline
   &$x$&$0$ & &$1$&\\
   $y$&&&&&\\\hline\hline
  $0$ &&$p(00|00) = 1-2p$&$ p(10|00) = p$  & $p(00|10) = 1-2p$&$ p(10|10) = p$     \\
   &&$p(01|00) = p$&$p(11|00) = 0$  & $p(01|10)=p$&$ p(11|10) = 0$  \\\hline
   $1$&&$p(00|01)=1-2p$&$ p(10|01)=p$  & $p(00|11)=1-p$&$ p(10|11)=0$   \\
  &&$p(01|01)=p$&$ p(11|01)=0$  & $p(01|11)=0$&$ p(11|11)=p$   \\\hline
\end{tabular}
\end{center}
 \caption{Correlations for the game with marginal probability p for the outcome 1}
\end{table}

The probability $p(00|00)$ is to be read as $p(a=0,b=0|x=0,y=0)$, and the probability $p(01|10)$ is to be read as $p(a=0,b=1|x=1,y=0)$, etc. (That is, I drop the commas for ease of reading; the first two slots in $p(--|--)$ before the conditionalization sign `$|$' represent the two possible outputs for Alice and Bob, respectively, and the second two slots after the conditionalization sign represent the two possible inputs for Alice and Bob, respectively.) Note that the sum of the probabilities in each square cell of the array in Table 1 is 1, and that the marginal probability of 0 for Alice and for Bob is obtained by adding the probabilities in the left column of each cell and the top row of each cell, respectively, and the marginal probability of 1 is obtained for Alice and for Bob by adding the probabilities in the right column of each cell and the bottom row of each cell, respectively. From Table 1, it is clear that the game defined by conditions 1, 2, 3 can be played with any value of $p$ in the range $0 \leq p \leq 1/2$.

The probability, for a particular strategy $S$, of winning the game with marginal $p$ and random inputs  is:
\begin{equation}
p_{S}\mbox{(win)} = \frac{1}{4}(p_{S}(a\cdot b = 0|00) + p_{S}(a\cdot b = 0|01) + p_{S}(a\cdot b = 0|10) + p_{S}(a\oplus b = 0|11))
\end{equation}
where the conditional probabilities are the probabilities of the outputs given the inputs, for the strategy $S$. 

We can express $\mbox{prob}_{S}\mbox{(win)}$ in terms of the marginal $p$ and the Clauser-Horne-Shimony-Holt (CHSH) correlation $K_{S}$ for the strategy $S$ (see \cite{CHSH}), where
\begin{equation} 
K_{S} = \langle 00\rangle_{S} + \langle 01\rangle_{S} + \langle 10\rangle_{S} - \langle 11\rangle_{S}
\end{equation}
Here $\langle xy\rangle_{S}$ is the expectation value, for the strategy $S$, of the product of the outputs for the input pair $x,y$, for the possible output values $\pm 1$ instead of 0 or 1:
\begin{equation}
\langle xy\rangle_{S} = p_{S}(1,1|xy) - p_{S}(1,-1|xy) - p_{S}(-1,1|xy) + p_{S}(-1,-1|xy)
\end{equation}
(inserting commas to separate Alice's output from Bob's output). So:
\begin{equation}
\langle xy\rangle_{S} = p_{S}(\mbox{outputs same}|xy) - p_{S}(\mbox{outputs different}|xy)
\end{equation}
and we can write:
\begin{eqnarray}
p_{S}(\mbox{outputs same}|xy) & = & \frac{1 +  \langle xy\rangle_{S}}{2} \\
p_{S}(\mbox{outputs different}|xy) & = & \frac{1-\langle xy\rangle_{S}}{2}
\end{eqnarray}

Since `$a\cdot b = 0$' is equivalent to `outputs different or 00' and `$a\oplus b = 0$' is equivalent to `outputs same':
\begin{eqnarray}
p_{S}\mbox{(win)} & = &  \frac{1}{4}(p_{S}(\mbox{outputs different}|00) + p_{S}(\mbox{outputs different}|01) \nonumber \\
& & + p_{S}(\mbox{outputs different}|10) + p_{S}(\mbox{outputs same}|11)) \nonumber \\
& & + \frac{1}{4}(p_{S}(00|00) + p_{S}(00|01) + p_{S}(00|10))
\end{eqnarray}
The probabilities of `outputs same' and `outputs different' are unchanged by the change of units to $\pm 1$ instead of 0 or 1 for the outputs, so:
\begin{equation}
p_{S}\mbox{(win)} =  \frac{1}{2} - \frac{K_{S}}{8} + \frac{3(1-2p)}{4}
\end{equation}

We consider the game under the assumption that the players have access to various sorts of resources: classical, quantum, or superquantum. To begin with, we assume that after Alice and Bob are separated, they are allowed to take with them any classical resources they like, such as any notes about the strategy, shared lists of random numbers, calculators or classical computers, classical measuring instruments, etc., but no quantum resources, such as entangled quantum states or quantum computers, or quantum measurement devices, and no hypothetical superquantum resources such as PR-boxes (see below).

Bell's locality argument \cite{BellEPR} in the CHSH version shows that if Alice and Bob are limited to classical resources, i.e., if they are required to reproduce the correlations on the basis of shared randomness or common causes established before they separate (after which no communication is allowed), then $|K_{C}| \leq 2$.  So a winning classical strategy $S = C$ (i.e., $p_{C}\mbox{(win)} = 1$) is impossible if $3(1-2p)/4 < 1/4$, i.e, if  $p > 1/3$.

In fact, there is a winning classical strategy for the game with $p \leq 1/3$. For $p = 1/3$, Alice and Bob generate a random sequence of digits 1, 2, 3, where each digit occurs with probability 1/3 in the sequence. They write down this sequence to as many digits as they like (at least as many as the rounds of the game) and they each take a copy of the list with them when they are separated at the start of the game. They associate the digits with the deterministic states in Tables 2, 3, 4 below (where the correlations are represented in abbreviated form, but  are to be read as in Table 1), and they take copies of these tables with them as well.

\begin{table}[h!]
\begin{center}
\begin{tabular}{|ll||ll|ll|} \hline
   &$x$&$0$ & &$1$&\\
   $y$&&&&&\\\hline\hline
  $0$ &&$1$&$0$  & $0$&$1$     \\
   &&$0$&$0$  & $0$&$0$  \\\hline
   $1$&&$0$&$0$  & $0$&$0$   \\
  &&$1$&$0$&$0$&$1$   \\\hline
\end{tabular}
\end{center}
 \caption{Deterministic state for random digit 1}
\end{table}

\begin{table}[h!]
\begin{center}
\begin{tabular}{|ll||ll|ll|} \hline
   &$x$&$0$ & &$1$&\\
   $y$&&&&&\\\hline\hline
  $0$ &&$0$&$1$  & $1$&$0$     \\
   &&$0$&$0$  & $0$&$0$  \\\hline
   $1$&&$0$&$1$  & $1$&$0$   \\
  &&$0$&$0$&$0$&$0$   \\\hline
\end{tabular}
\end{center}
 \caption{Deterministic state for random digit 2}
\end{table}

\begin{table}[h!]
\begin{center}
\begin{tabular}{|ll||ll|ll|} \hline
   &$x$&$0$ & &$1$&\\
   $y$&&&&&\\\hline\hline
  $0$ &&$0$&$0$  & $0$&$0$     \\
   &&$1$&$0$  & $1$&$0$  \\\hline
   $1$&&$1$&$0$  & $1$&$0$   \\
  &&$0$&$0$&$0$&$0$   \\\hline
\end{tabular}
\end{center}
 \caption{Deterministic state for random digit 3}
\end{table}

At each round of the game, they consult the random sequence, beginning with the first random digit for the first round, and they move sequentially through the random digits for subsequent rounds. They respond, for any given input, in terms of the entry in the appropriate box. For example, suppose the random digit in the sequence for a particular round is 2 and the input for Alice is $x=0$ and the input for Bob is $y=1$. The appropriate cell for these inputs in the table for random digit 2 (Table 3) is the bottom left cell. In this case, since the entries are all 0 except the entry $p(a=1, b=0|x=0, y=1) = p(10|01) = 1$,  Alice responds with the output 1 and Bob responds with the output 0. It is easy to see that this is a winning strategy by adding the corresponding entries in Tables 2, 3, 4 with weights 1/3---this produces Table 1. 

The three arrays in Tables 2, 3, 4 represent three deterministic states, each of which defines a definite response for each of the four possible combinations of inputs for Alice and Bob: 00, 01, 10, 11. The Table for random digit 1 corresponds to the deterministic state in which Alice outputs 1 if and only  if her input is 1, and Bob outputs 1 if and only if his input is 1 (otherwise they output 0). The Table for random digit 2 corresponds to the deterministic state in which Alice outputs 1 if and only if her input is 0, and Bob outputs 0 for both inputs. The Table for random digit 3 corresponds to the deterministic state in which Bob outputs 1 if and only if his input is 0, and Alice outputs 0 for both inputs. Note that each of the three joint deterministic states can be expressed as a product of local deterministic states for Alice and Bob separately. These states are the common causes of the correlated responses. In this case, by exploiting the resource of `shared randomness,' where each element in the random sequence is associated with a deterministic state, Alice and Bob can perfectly simulate the correlations of the $p = 1/3$ game. ((For $p < 1/3$, Alice and Bob generate a random sequence of digits 1, 2, 3, 4,  where the digits 1, 2, 3 occur with probability $p$ in the sequence and the digit 4 occurs with probability  $1 - 3p$. The digit 4 is associated with the deterministic state that assigns the output 0 to any input.)

Classical correlations are common cause correlations, i.e., they can be simulated perfectly by shared randomness. For $p > 1/3$, the correlations are nonclassical: they cannot be simulated by shared randomness, i.e., they are not common cause correlations.  If Alice and Bob are allowed quantum resources and base their strategy on measurements on shared entangled states prepared before they separate, then the Tsirelson bound  \cite{Tsirelson1980} $|K_{Q}| \leq 2\sqrt{2}$ applies. It follows that a winning quantum strategy $S = Q$  is impossible if $p > (1+\sqrt{2})/6$. 

Consider the game for $p = 1/2$, where the correlations are as in Table 6. In this case---since $p(00)$ must be 0 for inputs different from 11 if $p = 1/2$---Alice and Bob are required to produce different responses for the three pairs of inputs 00, 01, 10, and the same response for the input pair 11, with marginal probabilities of 1/2.
\begin{table}[h!]
\begin{center}
\begin{tabular}{|ll||ll|ll|} \hline
   &$x$&$0$ & &$1$&\\
   $y$&&&&&\\\hline\hline
  $0$ &&$0$&$1/2$  & $0$&$1/2$     \\
   &&$1/2$&$0$  & $1/2$&$0$  \\\hline
   $1$&&$0$&$1/2$  & $1/2$&$0$   \\
  &&$1/2$&$0$&$0$&$1/2$   \\\hline
\end{tabular}
\end{center}
 \caption{Correlations for the $p=1/2$ game}
\end{table}

The probability of winning the $p = 1/2$ game for random inputs is:
\begin{equation}
p_{S}\mbox{(win)} =  \frac{1}{2} - \frac{K}{8} 
\end{equation}
Since a winning classical or quantum strategy is impossible, we can consider the optimal probability of winning the $p=1/2$ game with classical or quantum resources. The optimal classical strategy is obtained for $K_{C} = -2$:
\begin{equation}
p_{\mbox{optimal}\, C}\mbox{(win)} = \frac{1}{2} - \frac{K}{8} = \frac{1}{2} + \frac{2}{8} = \frac{3}{4}
\end{equation}

In fact, the obvious classical strategy for winning three out of four rounds in the $p=1/2$ game would be for Alice and Bob to prepare a random sequence of 0's and 1's and take copies of this sequence with them at the start of the game. They consult the random sequence in order as the rounds proceed. If the random digit is 1, Alice outputs 1 and Bob outputs 0, for any input. If the random digit is 0, Alice outputs 0 and Bob outputs 1 for any input. Then the marginal probabilities of 0 and 1 will be 1/2, and Alice and Bob will produce different outputs for all four pairs of inputs---which means that they will produce the correct response for 3/4 of the rounds, on average. (If Alice and Bob both output 1 if the random digit is 1, and both output 0 if the random digit is 0, for any input, the marginal probabilities of 0 and 1 will still be 1/2, and they will produce the correct response for 1/4 of the rounds, on average, corresponding to $K_{C} = 2$. If they respond randomly, they will produce each of the output pairs 00, 01, 10, 11 with probability 1/4, and so they will produce the correct response for 1/2 the rounds on average, corresponding to $K_{C} = 0$. For all possible classical strategies, the probability of winning the game lies between 1/4 and 3/4.)

The optimal quantum strategy is obtained for $K_{Q} = -2\sqrt{2}$:
\begin{equation}
p_{\mbox{optimal}\, Q}\mbox{(win)} = \frac{1}{2} - \frac{K}{8} = \frac{1}{2} + \frac{2\sqrt{2}}{8}  \approx .85
\end{equation}

The correlations for the $p=1/2$ game are the correlations of a Popescu-Rohrlich (PR) box \cite{PopescuRohrlich94}, which are usually represented as in Table 7.\footnote{If Alice's (or Bob's) outputs are flipped for every input, the correlations of Table 6 are transformed to those of Table 7.}
\begin{table}[h!]
\begin{center}
\begin{tabular}{|ll||ll|ll|} \hline
   &$x$&$0$ & &$1$&\\
   $y$&&&&&\\\hline\hline
  $0$ &&$1/2$&$0$  & $1/2$&$0$     \\
   &&$0$&$1/2$  & $0$&$1/2$  \\\hline
   $1$&&$1/2$&$0$  & $0$&$1/2$   \\
  &&$0$&$1/2$&$1/2$&$0$   \\\hline
\end{tabular}
\end{center}
 \caption{Correlations for the PR-box}
\end{table}
Popescu and Rohrlich introduced the PR-box as a hypothetical device or nonlocal information channel that is more nonlocal than quantum mechanics, and in fact maximally nonlocal, in the sense that the correlations between outputs of the box for given inputs maximally violate the Tsirelson bound:
\begin{equation}
|K_{PR}| = |\langle 00\rangle_{PR} + \langle 01\rangle_{PR}  +  \langle 10\rangle_{PR} - \langle 11\rangle_{PR}| = 4
\end{equation} 
(since each of the four expectation values lies between $-1$ and $+1$).   

The defining correlations of a PR-box are specified by the relation:
\begin{equation}
a\oplus b = x\cdot y \label{eqn:PRbox}
\end{equation}
with marginal probabilities equal to 1/2 for all inputs and all outputs, i.e., 
\begin{itemize}
\item[1$'$.] if the inputs $x, y$ are 00, 01, or 10, then the outputs are the same (i.e., 00 or 11) 
\item[2$'$.]  if the inputs $x, y$ are 11, then the outputs are different (i.e., 01 or 10)
\item[3$'$.] $p(a|x) = p(b|y) = 1/2$, for all $a, b, x, y \in \{0,1\}$
\end{itemize}

It follows that the `no signaling' constraint is satisfied. Just as we considered 2-person games under the assumption that the players have access to classical or quantum resources, we can consider 2-person games---purely hypothetically---in which the players are allowed to take PR-boxes with them after the start of the game. It is assumed that the  $x$-input and $a$-ouput of a PR-box can be separated from the $y$-input and $b$-output by any distance without altering the correlations. Evidently, if Alice and Bob share many PR-boxes, where Alice holds the $x, a$  side of each box and Bob holds the $y, b$ side of each box, they can win the $p = 1/2$ game.

From the perspective of nonlocal PR boxes and other nonclassical correlations, we see that quantum correlations are not particularly special. Indeed, classical correlations appear to be rather special. The convex set of classical probability distributions has the structure of a simplex. An $n$-simplex is a particular sort of convex  set: a convex polytope\footnote{a polytope in two dimensions is a polygon, in three dimensions a polyhedron. More precisely, a convex polytope $\mathcal{P}$ is the convex hull of $n$ points $p_{1}, p_{2}, \ldots, p_{n}$, i.e., the set $\mathcal{P} = \{\sum_{i=1}^{n}\lambda_{i}p_{i}: \lambda_{i} \geq 0\, \mbox{for all}\, $i$\, \mbox{and}\, \sum_{i=1}^{n}\lambda_{i} = 1\}$.} generated by $n+1$ vertices that are not confined to any $(n-1)$-dimensional subspace (e.g., a triangle as opposed to a rectangle). The simplest classical state space in this sense (where the points of the space represent probability distributions) is the 1-bit space (the 1-simplex), consisting of two pure or extremal deterministic states, $\mathbf{0} = \left ( \begin{array}{c}1 \\ 0 \end{array} \right )$ and $\mathbf{1} = \left ( \begin{array}{c}0 \\ 1 \end{array} \right )$, represented by the vertices of the simplex, with mixed states---convex combinations of pure states---represented by the line segment between the two vertices: $\mathbf{p} = p\,\mathbf{0} + (1-p)\,\mathbf{1}$, for $0 \leq p \leq 1$. A simplex has the property that a mixed state can be represented in one and only one way as a mixture of extremal states, the vertices of the simplex. \emph{No other state space has this feature:} if the state space is not a simplex, the representation of mixed states as convex combinations of extremal states is not unique. The state space of classical mechanics is an infinite-dimensional simplex, where the extremal states are all deterministic states, with enough structure to support  transformations acting on the vertices that include the canonical transformations generated by Hamiltonians. 

The space of `no signaling' bipartite probability distributions, with arbitrary inputs $x \in \{1, \ldots, n\}, y \in \{1, \ldots, n\}$ and binary outputs, 0 or 1, is a convex polytope that is not a simplex---the `no signaling' correlational polytope---with the vertices (in the case $n = 2$) representing generalized PR-boxes (which differ from the standard PR-box only with respect to permutations of the inputs and/or outputs), or deterministic boxes (deterministic states), or (in the case $n > 2$) combinations of these (where the probabilities for some pairs of inputs are those of a generalized PR-box, while for other pairs of inputs they are the probabilities of a deterministic box; see \cite{JonesMasanes2005}, \cite{BLMPP2005}, \cite{BarrettPironio2005}). Inside this polytope is the convex set of quantum correlations, which is not a polytope (the  simplest quantum system is the qubit, whose state space as a convex set is a sphere: the Bloch sphere), and inside the quantum convex set is the convex set of classical correlations, which has the rather special structure of a simplex, where the extremal states are all deterministic states.\footnote{Note that these extremal deterministic states of the classical simplex are also extremal states of the `no signaling' correlational polytope. But the `no signaling' correlational polytope has additional nondeterministic extremal states, like the states defined by PR-boxes.} 

The non-unique decomposition of mixtures into convex combinations of  extremal states underlies the impossibility of a universal cloning  machine that can copy the extremal states of any probability distribution, the monogamy of nonclassical correlations, and various other features like `remote steering' in Schr\"{o}dinger's sense (see \S 3), which turn out to be generic features of nonclassical (i.e., non-simplex) theories \cite{Barrett2007}, \cite{Masanes2006}.

 Suppose, for example, that Bob could copy his part of a PR-box, with correlations as in Table 8, so that each part reproduced the PR-correlations.\footnote{Think of a PR-box as a nonlocal bipartite state, like a maximally entangled quantum state, e.g., the singlet state for spin-1/2 particles, or any of the Bell states. Copying one side of a PR-box would be like copying one half of a Bell state.} Then Bob has two inputs, $y$ and $y'$. Suppose Bob sets:
\begin{eqnarray}
y & = & 0 \nonumber \\
y' & = & 1 \nonumber
\end{eqnarray}
The PR-box correlations require:
\begin{eqnarray}
a\oplus b & = & x\cdot y \nonumber \\
a\oplus b' & = & x\cdot y' \nonumber
\end{eqnarray}
It follows that:
\begin{equation}
(a\oplus b) \oplus (a\oplus b') = b \oplus b' = x\cdot (y \oplus y') = x
\end{equation}
So Bob could compute $x$, the value of Alice's input, from the Boolean sum of his two outputs: if his outputs take the same value, then Alice's input is 0; if they take opposite values, Alice's input is 1. If such a cloning device were possible,  Alice and Bob could use the combined PR-box and cloning device to signal instantaneously. Since we are assuming `no signaling,' such a device must be impossible. An analogous argument applies not only to the hypothetical correlations of nonlocal boxes such as the PR-box, but to quantum correlations, i.e., there can be no device that will copy one half of an entangled quantum state without allowing the possibility of instantaneous signaling. 

Similarly, nonclassical correlations are monogamous: the correlations of a PR-box, for example, can be shared by Alice and Bob, but not by Alice and Bob as well as Charles. If the correlations with Alice could be shared by Bob and Charles, then Bob and Charles could use their outputs to infer Alice's input, allowing instantaneous signaling between Alice and Bob-Charles. By contrast, there is no such constraint on classical correlations: Alice can happily share any classical correlations with Bob and also with Charles, David, \ldots without violating the `no signaling' constraint.

This is essentially because classical correlations between Alice and Bob can be reduced uniquely to a shared probability distribution over joint deterministic states that are also product deterministic states (local states) for Alice and Bob separately. Since the deterministic states can be reproduced by a set of instructions for Alice that relate inputs to outputs deterministically, and a separate and independent set of instructions for Bob that relate inputs to outputs deterministically, Alice and Bob can simulate any classical correlations. And since nothing prevents Bob, say, from sharing his local instructions with other parties as many times as he likes,  Alice can share these correlations with Bob as well as any number of other parties. 

For example, suppose the correlations are as in Table 2, so that Alice and Bob both output 0 when their inputs are both 0, both output 1 when their inputs are both 1, and output different outputs when their inputs are different. Alice and Bob achieve this correlation by simply following the separate deterministic rules:
\begin{eqnarray}
a = x \nonumber \\
b = y \nonumber
\end{eqnarray}
and clearly Alice and Charles can achieve the same correlation by following the deterministic rules:
\begin{eqnarray}
a = x \nonumber \\
c = z \nonumber
\end{eqnarray}
without violating the `no signaling' constraint.

\begin{figure}
\begin{picture}(100,150)(-65,0)
\scalebox{0.7}{\includegraphics{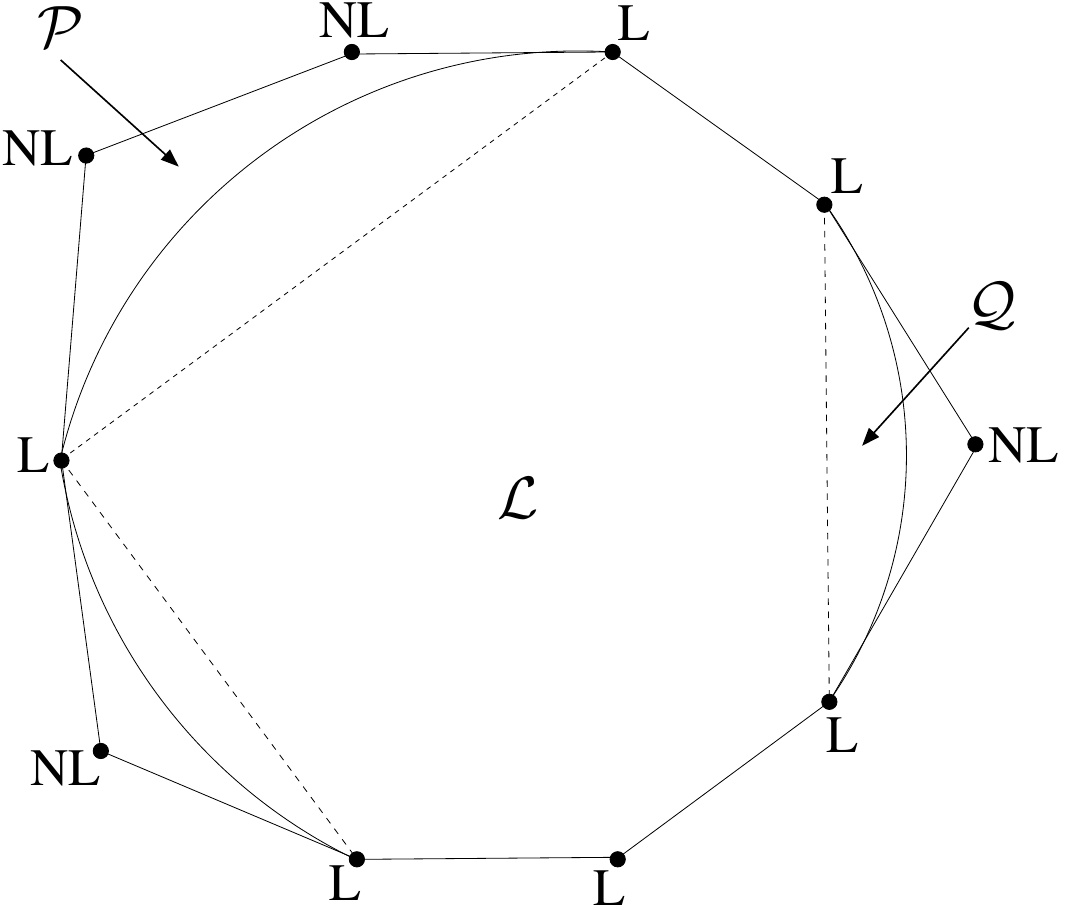}}
\end{picture}
\caption[]{A schematic
representation of the space of no-signaling correlations.  The
vertices are labelled L and NL for local and nonlocal. Bell inequalities characterize the facets represented by dashed lines. The set bounded by these is $\mathcal{L}$. The
region accessible to quantum mechanics is $\mathcal{Q}$. Superquantum correlations lie in region $\mathcal{P}$ outside the quantum region. From a diagram in \cite{BLMPP2005}.}
\end{figure}

To sum up this section: The class of `no signaling' theories includes classical theories, quantum theories, and superquantum theories. Classical theories are characterized as theories whose state spaces have the structure of a simplex. This guarantees a \emph{unique} decomposition of any mixed state into a convex combination of pure or extremal classical states, which are all deterministic states. Note that the lattice of subspaces of a simplex (the lattice of vertices, edges, and faces) is a Boolean algebra, with a 1-1 correspondence between the vertices, corresponding to the atoms of the Boolean algebra, and the facets (the $(n-1)$-dimensional faces), corresponding to the co-atoms. The classical simplex represents the classical state space regarded as a space of classical (multipartite) probability distributions; the associated Boolean algebra represents the classical event structure. The conceptually puzzling features of nonclassical `no signaling' theories---quantum and superquantum---can all be associated with state spaces that have the structure of a polytope whose vertices include the local deterministic extremal states of the classical simplex, as well as nonlocal nondeterministic extremal states (like PR-boxes) that lie outside the classical simplex (see Fig. 1).  Mixed states that lie outside the classical polytope decompose \emph{non-uniquely} into convex combinations of these extremal states. The non-unique decomposition of mixed states into convex combinations of pure states is a characteristic feature of nonclassical `no signaling' theories, including quantum theories.

In the following section, I take a closer look at quantum probabilities and quantum correlations, and in particular quantum conditionalization, which involves a loss of information that is a generic feature of conditionalization in nonclasscal `no signaling' theories.

\section{Conditionalizing Quantum Probabilities}

The event space of a classical system is represented by the Boolean algebra of subsets (technically, the Borel subsets) of the phase space of the system. The event space of a quantum system is represented by the closed subspaces of a Hilbert space, which form an infinite collection of intertwined Boolean algebras. Each Boolean algebra corresponds to a partition of the Hilbert space representing a collection of mutually exclusive and collectively exhaustive events. If $e$ and $f$ are atomic events, represented by 1-dimensional subspaces spanned by the vectors $\ket{e}, \ket{f}$, the probability of the event $f$ given the event $e$ is given by the Born rule:
\begin{equation}
\mbox{prob}(e,f)  = |\langle e|f\rangle|^{2} =  |\langle f|e\rangle|^{2} = \cos^{2}\theta_{ef}
\end{equation}
More generally, the probability of an event $a$ (not necessarily atomic) can be expressed as:
\begin{equation}
\mbox{prob}_{\rho}(a) = \trace{\rho P_{a}} \label{eqn:qtracerule}
\end{equation}
where $P_{a}$ is the projection operator onto the subspace representing the event $a$ and $\rho$ is a density operator representing a pure state ($\rho = P_{e}$, for some atomic event $e$) or a mixed state ($\rho = \sum_{i}w_{i}P_{e_{i}}$).  Gleason's theorem \cite{Gleason} shows that  this representation of quantum probabilities is unique in a Hilbert space $\hil{H}$ of dimension greater than 2. 

Conditionalization on the occurrence of an event $a$, in the sense of a
minimal revision---consistent with the subspace structure of Hilbert
space---of the probabilistic information encoded in a quantum state given by a density operator $\rho$, is given by the von Neumann-L\"{u}ders rule (the `projection postulate' if  $\rho$ is a pure state):
\begin{equation}
\rho \rightarrow \rho_{a} \equiv \frac{P_{a}\rho P_{a}}{\trace{(P_{a} \rho P_{a})}} \label{eqn:luders}
\end{equation}
where $P_{a}$ is the projection operator onto the subspace representing the
event $a$. Here $\rho_{a}$ is the conditionalized density operator,
conditional on the event $a$, and the normalizing factor $\trace{(P_{a} \rho P_{a})} =\trace{(\rho P_{a})}$
is the probability assigned to the event $a$ by the state $\rho$. So the conditional probability of an event $b$, given an event $a$, is:
\begin{equation}
\mbox{prob}_{\rho}(b|a) =  \trace{(\rho_{a}P_{b})} \label{eq:qcondprob}
\end{equation}

The transition (\ref{eqn:luders}) is  just  a generalization, in the non-Boolean quantum event space, of the classical Bayesian rule for updating an initial probability distribution on new information.\footnote{The analysis can be extended to the general case of measurements represented by positive operator valued measures (POVM's). See \cite{Henderson}. A general measurement represented by a POVM  on a system $S \in \hil{H}_{S}$ is equivalent to a projection-valued measurement on a larger Hilbert space: specifically, a projective measurement on an ancilla system $E \in \hil{H}_{E}$ suitably entangled with $S$. An analogous equivalence holds for classical systems. For an account of such general measurements, see the section on measurement in \cite{NielsenChuang} or  \cite{BubQIC}.} 
To see this, consider a countable classical probability space $(X,\hil{F}, \mu)$, with atomic or elementary events  $x_{1}, x_{2}, \ldots$ associated with singleton subsets $X_{1}, X_{2}, \ldots$ and characteristic functions $\chi_{1}, \chi_{2}, \ldots$ The atomic characteristic functions define deterministic states that assign probability 1 to the corresponding atomic event and probability 0 to all other events. Denote non-atomic events by $a, b, \ldots$ and the characteristic functions associated with the corresponding subsets $X_{a}, X_{b}, \ldots$ by $\chi_{a}, \chi_{b}, \ldots$ 
  
Since any classical probability measure $\mu$ can be expressed uniquely as a mixture of deterministic (extremal) states with probabilities $p_{i}$, it is possible to associate a unique `density operator' $\rho = \sum_{i}p_{i}\chi_{i}$ (where $\sum_{i}p_{i} = 1$, $p_{i} \geq 0$, for all $i$) with  $\mu$,  in terms of which the probability of an event $a$ can be represented as:
\begin{eqnarray}
\mbox{prob}_{\mu}(a) & = & \mu(X_{a}) \nonumber \\
& = &  \sum_{j}\left(\sum_{i}p_{i}\chi_{i}(x_{j})\right)\chi_{a}(x_{j}) \nonumber \\
& = & \sum_{j}\rho(x_{j})\chi_{a}(x_{j}) 
\end{eqnarray}
Writing $\mbox{prob}_{\rho}(a)$ for $\mbox{prob}_{\mu}(a)$, we have: 
\begin{equation}
\mbox{prob}_{\rho}(a)  =  \sum \rho \chi_{a} \label{eqn:ctracerule}
\end{equation}
where a summation sign without an index is understood as summing over all the atomic events. Equation (\ref{eqn:ctracerule}) is the classical analogue of (\ref{eqn:qtracerule}). Note that the trace of an operator $O$ is just the sum of the eigenvalues of $O$, i.e., the sum of the possible values of $O$ at each atom in the Boolean subalgebra defined by $O$.

The conditional probability of an event $b$, given an event $a$: 
\begin{equation}
\mbox{prob}_{\mu}(b|a)= \frac{\mu(X_{a}\cap X_{b})}{\mu(X_{a})}
\end{equation}
can be represented in terms of the density operator $\rho$,  as:
\begin{eqnarray}
\mbox{prob}_{\rho}(b|a) & = &  \frac{\sum_{j} \rho(x_{j})\chi_{a}(x_{j})\chi_{b}(x_{j})}{\sum_{j} \rho(x_{j})\chi_{a}(x_{j})} \nonumber \\
& = & \frac{\sum \rho \chi_{a}\chi_{b}}{\sum \rho \chi_{a}}
\end{eqnarray}

The transition
\begin{equation}
\mu \rightarrow \mu_{a},
\end{equation}
where $\mu_{a}$ is defined for any event $b$ by:
\begin{equation}
\mu_{a}(X_{b}) \equiv \frac{\mu(X_{a}\cap X_{b})}{\mu(X_{a})}
\end{equation}
represents  the classical Bayesian rule for updating an initial probability distribution on new information $a$. It can be justified in terms of coherence constraints by a Dutch book argument. The rule can be represented in terms of the density operator $\rho$ as the transition:
\begin{equation}
\rho  \rightarrow \rho_{a} \equiv \frac{\rho \chi_{a}}{\sum \rho \chi_{a}}
\end{equation}
or, equivalently, in the symmetrized form:
\begin{equation}
\rho  \rightarrow \rho_{a} \equiv \frac{\chi_{a}\rho \chi_{a}}{\sum \chi_{a}\rho \chi_{a}} \label{eq:Bayesclass}
\end{equation}
so that 
\begin{equation}
\mbox{prob}_{\rho}(b|a) = \sum \rho_{a} \chi_{b} \label{eq:ccondprob}
\end{equation}

We see that the von Neumann-L\"{u}ders rule (\ref{eqn:luders}) is the quantum analogue of the classical Bayesian updating rule  (\ref{eq:Bayesclass}), and (\ref{eq:qcondprob}) is the quantum analogue of (\ref{eq:ccondprob}).

If we consider a pair of correlated systems, $\mathbf{A}$ and $\mathbf{B}$, then
conditionalization on an $\mathbf{A}$-event, for the probabilistic information encoded
in the density operator $\rho_{\mathbf{B}}$ representing the probabilities of events
at the remote system $\mathbf{B}$, will always be an updating, in the sense of a
refinement of the information available at system $\mathbf{A}$ about system $\mathbf{B}$, i.e, the selection of a particular alternative (depending on the $\mathbf{A}$-event) in a particular set of mutually exclusive and collectively exhaustive alternatives (depending on the \emph{type} of $\mathbf{A}$-event, i.e., the observable measured at $\mathbf{A}$).

For example, suppose the system $\mathbf{A}$ is associated with a 3-dimensional Hilbert
space $\mathcal{H}_{\mathbf{A}}$ and the system $\mathbf{B}$ is associated with a
2-dimensional Hilbert space $\mathcal{H}_{\mathbf{B}}$.\footnote{I use boldfaced letters to denote the two systems, $\mathbf{A}$ and $\mathbf{B}$, and italic symbols to denote observables, e.g., $A, A'$ for $\mathbf{A}$-observables and $B, B'$ for $\mathbf{B}$-observables.} Suppose the
composite system $\mathbf{AB}$ is in an entangled state: 
\begin{eqnarray}
| \psi^{\mathbf{AB}} \rangle & = & \frac{1}{\sqrt{3}}(| a_{1} \rangle| b_{1} \rangle
+ | a_{2} \rangle| c \rangle + | a_{3} \rangle| d \rangle)  \notag \\
& = & \frac{1}{\sqrt{3}}(| a^{\prime}_{1} \rangle| b_{2} \rangle + |
a^{\prime}_{2} \rangle| e \rangle + | a^{\prime}_{3} \rangle| f \rangle)
\end{eqnarray}
where $| a_{1} \rangle,| a_{2} \rangle,| a_{3} \rangle$ and $|
a^{\prime}_{1} \rangle,| a^{\prime}_{2} \rangle,| a^{\prime}_{3} \rangle$
are two orthonormal bases in $\mbox{$\mathcal{H}$}_{\mathbf{A}}$ and $| b_{1}
\rangle, | b_{2} \rangle$ is an orthonormal basis in $\mbox{$\mathcal{H}$}
_{\mathbf{B}}$. The triple $| b_{1} \rangle,| c \rangle,| d \rangle$ and the triple $
| b_{2} \rangle,| e \rangle,| f \rangle$ are non-orthogonal triples of
vectors in $\mbox{$\mathcal{H}$}_{\mathbf{B}}$, where the vectors in each triple are separated by an angle $2\pi/3$. \footnote{For a precise
specification of these vectors, see Bub \cite{Bub2007}.} The reduced state of $\mathbf{B}$ (obtained
by tracing over $\mbox{$\mathcal{H}$}_{\mathbf{A}}$, i.e., $\rho_{\mathbf{B}} = \ptrace{\mathbf{A}}{\rho}$) is the completely mixed state $
\rho_{\mathbf{B}} = \frac{1}{2}I_{\mathbf{B}}$: 
\begin{equation}
\frac{1}{3}| b_{1} \rangle\langle b_{1} | + \frac{1}{3}| c \rangle\langle c
| + \frac{1}{3}| d \rangle\langle d | = \frac{1}{3}| b_{2} \rangle\langle
b_{2} | + \frac{1}{3}| e \rangle\langle e | + \frac{1}{3}| f \rangle\langle
f | = \frac{I_{\mathbf{B}}}{2}
\end{equation}

Conditionalizing on one of the eigenvalues $a_{1},a_{2},a_{3}$ or $
a_{1}^{\prime },a_{2}^{\prime },a_{3}^{\prime }$ of an $\mathbf{A}$-observable $A$ or $
A^{\prime }$ via (\ref{eqn:luders}), i.e., on the occurrence of an event
corresponding to $A$ taking the value $a_{i}$ or $A^{\prime }$ taking the
value $a_{i}^{\prime }$ for some $i$, changes the density operator $\rho _{\mathbf{B}}
$ of the remote system $\mathbf{B}$ to one of the states $|b_{1}\rangle ,|c\rangle
,|d\rangle $ or to one of the states $|b_{2}\rangle ,|e\rangle ,|f\rangle $.
Since the mixed state $\rho _{\mathbf{B}}=\frac{1}{2}I_{\mathbf{B}}$ can be decomposed as an
equal weight mixture of $|b_{1}\rangle ,|c\rangle ,|d\rangle $ or as an
equal weight mixture of $|b_{2}\rangle ,|e\rangle ,|f\rangle $, the change
in the state of $\mathbf{B}$ is an updating, in the sense of a refinement of the
information about $\mathbf{B}$ encoded in the state $|\psi ^{\mathbf{AB}}\rangle $, taking into
account the new information $a_{i}$ or $a_{i}^{\prime }$. In fact, the mixed
state $\rho _{\mathbf{B}}=\frac{1}{2}I_{\mathbf{B}}$ corresponds to an infinite variety of
mixtures of pure states in $\mbox{$\mathcal{H}$}_{\mathbf{B}}$ (not necessarily equal
weight mixtures, of course). The effect at the remote system $\mathbf{B}$ of
conditionalization on any event at $\mathbf{A}$ will always be an updating, in the
sense of a refinement, with respect to one the these mixtures.\footnote{
Fuchs makes a similar point in \cite{Fuchs2002b}.} This is the content of
the Hughston-Jozsa-Wootters theorem \cite{HJW}.  

Schr\"{o}dinger \cite[p. 556]{Schr1}
found this objectionable as a sort of  remote `steering,' in the sense that Alice at $\mathbf{A}$ can choose to measure $A$ or $A'$ and by doing so  `steer' $\mathbf{B}$ into a mixture of pure states $|b_{1}\rangle, |c\rangle, |d\rangle $ or into a mixture of pure states $|b_{2}\rangle, |e\rangle, |f\rangle $, at will. Remote steering is exploited in the phenomena of quantum teleportation and quantum
dense coding, and  underlies the
impossibility of unconditionally secure quantum bit commitment (see Bub \cite{BubQIC}
for a discussion). Nevertheless, nothing changes at $\mathbf{B}$ as a consequence of Alice's measurement at $\mathbf{A}$. The effect of conditionalization at a remote system (the system that is not
directly involved in the conditionalizing event) is  consistent with the
`no signaling' constraint (\ref{eqn:nosignal1}), (\ref{eqn:nosignal2}).  What is new here, relative to classical correlations, is the possibility of simultaneously correlating the values of different noncommuting $\mathbf{A}$-observables with the values of different noncommuting $\mathbf{B}$-observables  in an entangled state, even though the correlated values cannot all be definite simultaneously (i.e., even though the events corresponding to the observables taking a selection of the correlated values, one possible pair of values for each pair of correlated observables, cannot all occur simutaneously). What Alice is able to choose, by her choice of measurement, is just one of these correlated pairs. Then the change in probabilities at the remote system  $\mathbf{B}$ when Alice conditionalizes on the value of the chosen $\mathbf{A}$-observable is simply an updating in the sense of a refinement of the prior information about $\mathbf{B}$
expressed in terms of the correlation between the chosen $\mathbf{A}$-observable and the correlated $\mathbf{B}$-observable,
as encoded in the entangled state $| \psi^{\mathbf{AB}} \rangle$. If this were not the case, i.e., if averaging over the possible outcomes of an $\mathbf{A}$-measurement yielded marginal probabilities at $\mathbf{B}$ that depended on the observable measured at $\mathbf{A}$, then the reduced state $\rho _{\mathbf{B}}$, obtained by tracing over $\mathcal{H}_{\mathbf{A}}$, would not be independent of the $\mathbf{A}$-basis chosen,\footnote{In that case, Hilbert space would not be an appropriate representation space for quantum states and quantum events.} and  instantaneous signaling between $\mathbf{A}$ and $\mathbf{B}$ would be possible. The
occurrence of a particular sort of event at $\mathbf{A}$---corresponding to a definite value
for the observable $A$ as opposed to a definite value for some other observable $A^{\prime}$---would produce a detectable change in the $\mathbf{B}$-probabilities, and so Alice at
$\mathbf{A}$ could signal instantaneously to Bob at $\mathbf{B}$ merely by choosing to perform a particular $\mathbf{A}$-measurement, $A$ or $A'$, and gaining a specific sort of information about $\mathbf{A}$ (the value
of $A$ or the value of $A^{\prime}$).

To avoid violating the `no signaling' constraint,  it must be impossible to construct a cloning machine that will clone the
extremal states of a quantum probability distribution defined by an arbitrary density operator. 
For suppose a universal cloning machine were possible. Then such a device could
copy any state in the non-orthogonal triple $| b_{1} \rangle, | c \rangle, | d
\rangle$ as well as any state in the non-orthogonal triple $| b_{2} \rangle, | e
\rangle, | f \rangle$. It would then be possible for Alice at $\mathbf{A}$ to signal to
Bob at $\mathbf{B}$. If Alice obtained the information given by an eigenvalue $a_{i}$ of 
$A$ or $a^{\prime}_{i}$ of $A^{\prime}$, and Bob were to input the system $\mathbf{B}$ into
the cloning device $n$ times, he would obtain one of the states $| b_{1}
\rangle^{\otimes n}, | c \rangle^{\otimes n}, | d \rangle^{\otimes n}$ or
one of the states $| b_{2} \rangle^{\otimes n}, | e \rangle^{\otimes n}, | f
\rangle^{\otimes n}$, depending on the nature of Alice's information. Since
these states tend to mutual orthogonality in $\otimes^{n}\mbox{$%
\mathcal{H_{\mathbf{B}}}$}$ as $n \rightarrow \infty$, they are distinguishable in
the limit. So, even for finite $n$, Bob would in principle be able to obtain
some information instantaneously about a remote event.

More fundamentally, the existence of a universal cloning machine for quantum pure states is
inconsistent with the interpretation of Hilbert space as  the kinematic framework for an indeterministic physics (see \S 5). For such a device
would be able to distinguish the equivalent mixtures of non-orthogonal states
represented by the same density operator $\rho_{\mathbf{B}} = \frac{1}{2}I_{\mathbf{B}}$. If a
quantum state prepared as an equal weight mixture of the states $| b_{1}
\rangle, | c \rangle, | d \rangle$ could be distinguished from a state
prepared as an equal weight mixture of the states $| b_{2} \rangle, | e
\rangle, | f \rangle$, the representation of quantum states by Hilbert space density
operators would be incomplete.

Now consider the effect of conditionalization on the state of $\mathbf{A}$. The state
of $\mathbf{AB}$ can be expressed as the biorthogonal (Schmidt) decomposition: 
\begin{equation}
| \psi^{\mathbf{AB}} \rangle = \frac{1}{\sqrt{2}} (| g \rangle| b_{1} \rangle + | h
\rangle| b_{2} \rangle)
\end{equation}
where 
\begin{eqnarray}
| g \rangle & = & \frac{2| a_{1} \rangle - | a_{2} \rangle -| a_{3} \rangle}{
\sqrt{6}} \\
| h \rangle & = & \frac{| a_{2} \rangle - | a_{3} \rangle}{\sqrt{2}}
\end{eqnarray}
The density operator $\rho_{\mathbf{A}}$, obtained by tracing $| \psi^{\mathbf{A}} \rangle$
over $\mathbf{B}$, is: 
\begin{equation}
\rho_{\mathbf{A}} = \frac{1}{2}| g \rangle\langle g | + \frac{1}{2}| h \rangle\langle
h |
\end{equation}
which has support on a 2-dimensional subspace in the 3-dimensional Hilbert
space $\mbox{$\mathcal{H}$}_{\mathbf{A}}$: the plane spanned by $| g \rangle$ and $|
h \rangle$ (in fact, $\rho_{\mathbf{A}} = \frac{1}{2}P_{\mathbf{A}}$, where $P_{\mathbf{A}}$ is the
projection operator onto the plane). Conditionalizing on a value of $A$ or $
A^{\prime}$ yields a state that has a component outside this plane. So the
state change on conditionalization cannot be interpreted as an updating of
information in the sense of a refinement, i.e., as the selection of a
particular alternative among a set of mutually exclusive and collectively
exhaustive alternatives represented by the state $\rho_{\mathbf{A}}$.

This is the notorious `irreducible and uncontrollable disturbance' arising
in the act of recording the occurrence of an event in a quantum measurement process that
underlies the so-called measurement problem of quantum mechanics: the loss of some of the information
encoded in the original state (in the above example, the probability of the
$\mathbf{A}$-event represented by the projection operator onto the 2-dimensional
subspace $P_{\mathbf{A}}$ is no longer 1, after the registration of the new
information about the observable $A$ or $A^{\prime}$). Note, though, that a similar loss of information is a generic feature of conditionalization in nonclassical `no signaling' theories---certainly in all nonclassical theories in which the states are completely specified by the probabilities of the measurement outcomes of a finite, informationally complete (or `fiducial') set of observables (see \cite{BubPitowsky2010}, \cite{Barrett2007}). This is the case for a large class of nonclassical theories, including quantum mechanics. In the case of a qubit, for example, the probabilities for spin `up' and
spin `down' in three orthogonal directions suffice to define a direction on
the Bloch sphere and hence to determine the state, so the spin observables $
\sigma_{x}, \sigma_{y}, \sigma_{z}$ form an informationally complete set.\footnote{Note that an informationally complete set is not unique.}
(For a classical system, an
informationally complete set is given by  a single observable, with $n$
possible outcomes, for some $n$.)

Suppose $\mbox{$\mathcal{F}$} = \{Q,Q',\ldots\}$ is an informationally complete
set of observables with $n$ possible values.\footnote{The following argument is reproduced from \cite{BubPitowsky2010}. See the Appendix: The Information Loss Theorem.}  A state $
\rho$ assigns a probability distribution to every outcome of any measurement
of an obervable in $\mbox{$\mathcal{F}$}$. Measuring $Q$ yields one of the
outcomes $q_{1}, q_{2}, \ldots$ with a probability distribution $
p_{\rho}(q_{1}|Q), p_{\rho}(q_{2}|Q), \ldots$. Similarly, measuring $Q'$
yields one of the outcomes $q'_{1}, q'_{2}, \ldots$ with a probability
distribution $p_{\rho}(q'_{1}|Q'), p_{\rho}(q'_{2}|Q'), \ldots$, and so on. If $
\mbox{$\mathcal{F}$}$ is informationally complete, the finite set of
probabilities completely characterizes $\rho$.

Assuming that all measurement outcomes are independent and ignoring any
algebraic relations among elements of $\mbox{$\mathcal{F}$}$, a classical
probability measure on a classical (Kolmogorov) probability space can be
constructed from these probabilities: 
\begin{equation}
p_{\rho}(q,q,\ldots|Q,Q',\ldots) = p_{\rho}(q|Q)p_{\rho}(q'|Q')\ldots
\end{equation}
Note that the probability space is finite since $
\mbox{$\mathcal{F}$}$ is finite. (The number of atoms in the probability
space is at most $n^{|\mathcal{F}|}$.) The state $\rho$ can be reconstructed from 
$p_{\rho}$.

Now, suppose we are given an unknown arbitrary state $\rho$. Suppose it is possible to measure $Q,Q',\ldots$ sufficiently many times to generate the classical probability measure $P_{\rho}$, to as good an approximation as required, without  destroying $\rho$. From $P_{\rho}$ we could then construct a copy  of $\rho$:
\begin{equation}
\rho \stackrel{\mbox{\scriptsize measure}}{\longrightarrow} P_{\rho} \stackrel{\mbox{\scriptsize prepare}}{\longrightarrow} \rho 
\end{equation}

This procedure defines a universal cloning machine, which is impossible in a nonclassical `no signaling' theory. It follows that it must be impossible to generate the classical probability measure $P_{\rho}$ from $\rho$ in the manner described (which is the case in quantum mechanics if we have only one copy of $\rho$, or too few copies of $\rho$), or else, if we can generate $P_{\rho}$ from $\rho$, the original state $\rho$ must be changed irreversibly by the process of extracting the information to generate $P_{\rho}$ (if not, the change in $\rho$ could be reversed dynamically and cloning would be possible):
\begin{equation}
\xcancel{\rho} \stackrel{\mbox{\scriptsize measure}}{\longrightarrow} P_{\rho} \stackrel{\mbox{\scriptsize prepare}}{\longrightarrow} \rho \label{eqn:dist}
\end{equation}

So extracting information from a nonclassical `no signal' information source given by a  state $\rho$, sufficient to generate the probabilities of an informationally
complete set of observables, is either impossible or necessarily changes the
state $\rho$ irreversibly, i.e., there must be information loss in the
extraction of such information. Hence, no  complete dynamical account of the state transition in a
measurement process is possible in a nonclassical `no signaling' theory (because any measurement can be part of an informationally complete set, so any measurement must lead to an irreversible change in the state of the measured system).

Since cloning is impossible for an arbitrary quantum pure state,  there can be no measurement device that functions
dynamically in such a way as to identify with certainty an arbitrary quantum pure state, without altering the source
irreversibly or `uncontrollably,' to use Bohr's term---no device can
distinguish a given quantum pure state from every other possible pure state by undergoing a
dynamical (unitary) transformation that results in a state that represents a
distinguishable record of the output, without an irreversible transformation
of the state.

The preceding remarks apply to the class of theories considered. It might seem that they are directly contradicted by Bohm's theory, for example. In Bohm's theory, a collection of particles distributed in space is represented by a point in configuration space. The motion of the particles is guided by the quantum state represented as a function in configuration space, which evolves unitarily (i.e., in accordance with Schr\"{o}dinger's equation of motion). The theory `saves the appearances' by correlating particle positions with phenomena, on the assumption that the distribution of particle positions has reached the equilibrium Born distribution (which, once achieved, is shown to remain stable).	

In one sense, Bohm's theory is classical, with position in configuration space as the single informationally complete observable. Note, though, that the probabilities defined by a given quantum state for different observables cannot simply be derived from the distribution of particle positions---the quantum observables aren't functions of position. Rather, the quantum probabilities for the possible outcomes of a measurement of an observable are generated as the probabilities of particle trajectories, guided by the evolution of the quantum state for the measurement interaction, which depends on the observable measured. From the perspective of the theory, observables other than position do not represent physical quantities of the measured system, and what we refer to as  `measuring' an observable is not a measurement in the usual sense, but a particular sort of evolution of the wave function, manifested in the distribution of particle trajectories. 

For example, the momentum of a Bohmian particle is the rate of change of position, but the expectation value of the momentum observable in a quantum ensemble defined by the equilibrium Born distribution is not derived by averaging over the particle momenta. Bohm \cite[p 387]{Bohm2} gives an example of a free particle in a box of length $L$ with perfectly reflecting walls. Because the wave function is real, the particle is at rest. The kinetic energy of the particle is $E = P^{2}/2m = (nh/L)^{2}/2m$. Bohm asks: how can a particle with high energy be at rest in the empty space of the box? The solution to the puzzle is that a measurement of the particle's momentum involves a change in the wave function, which plays the role of a guiding field for the particle's motion, in such a way that the \emph{measured} momentum values turn out to be $\pm nh/L$ with equal probability. Bohm comments \cite[pp. 386--387]{Bohm2}:
\begin{quotation}
This means that the measurement of an `observable' is not really a measurement of any physical property belonging to the observed system alone. Instead, the value of an `observable' measures only an incompletely predictable and controllable potentiality belonging just as much to the measuring apparatus as to the observed system itself.
\ldots
We conclude then that this measurement of the momentum `observable' leads to the same result as is predicted in the usual interpretation. However, the actual particle momentum existing before the measurement took place is quite different from the numerical value obtained for the momentum `observable,' which, in the usual interpretation, is called the 'momentum.'
\end{quotation}

In another sense, Bohm's theory, in its general form, is not at all classical---it is not even in the class of `no signaling' theories. If Alice and Bob share entangled pairs of particles, and Bob, remote from Alice, measures an observable $B$ on his particle, the wave function evolves to a form characteristic of a $B$-measurement, which instantaneously affects the motion of Alice's particles in a particular way. So Alice can tell whether Bob measured $B$ or $B'$ by looking at the statistical behavior of her particle trajectories. There is no difference in the statistics only if the original distribution is the equilibrium distribution---in that case, the phenomena are just as predicted by quantum mechanics, and there is no detectable difference between Bohm's theory and standard quantum mechanics. 

Bohm's theory is ingenious and is certainly not ruled out by the preceding remarks. For all we know, Bohm's theory might be true. But one might say the same for Lorentz's theory in relation to special relativity, insofar as it `saves the appearances.' Lorentz's theory provides a dynamical explanation for phenomena, such as length contraction, that are explained kinematically in special relativity in terms of the structure of Minkowski space. The theory does this at the expense of introducing motions relative to the ether, which are in principle unmeasurable, given the equations of motion of the theory. Similarly, Bohm's theory provides a dynamical explanation of quantum phenomena, such as the loss of information on measurement, which is explained kinematically in quantum mechanics in terms of the structure of Hilbert space, at the expense of introducing  the positions of the Bohmian particles, which are in principle unmeasurable more precisely than the Born distribution in the equilibrium theory, given the equations of motion of the particles.

Ultimately, the question is whether it is more fruitful in terms of advancing our understanding to consider quantum mechanics as a member of the class of `no signaling' theories, where the observables of the theory represent physical quantities and the states define probability distributions, or whether we should think of quantum mechanics as an equilibrium version of a theory that violates the `no signaling' constraint in recovering the quantum statistics for the outcomes of what we regard as measurements.

\section{Two Measurement Problems}

The discussion in the previous sections concerned classical and nonclassical probabilities, in particular the peculiar probabilistic correlations between separated quantum systems. The purpose was to set the stage for a formulation and resolution of the fundamental interpretative problem of quantum mechanics: how to connect \emph{probability} with \emph{truth} in quantum world, i.e., how to relate quantum probabilities to the objective occurrence and non-occurrence of events.  

The problem is usually formulated as the measurement problem of quantum mechanics, along the following lines: Suppose a system, $\mathbf{S}$, is in one of two orthogonal states, $\ket{0}$ or $\ket{1}$, the eigenstates of an observable $Q$ with two eigenvalues, 0 or 1. A suitable (ideal) measuring instrument for $Q$ would be a system, $\mathbf{M}$, in a  neutral `ready' state, which could interact with $\mathbf{S}$ in such a way as to lead to the following dynamical evolution:
\begin{eqnarray}
\ket{0}\ket{\mbox{ready}} \rightarrow \ket{0}\ket{'0'} \\
\ket{1}\ket{\mbox{ready}} \rightarrow \ket{1}\ket{'1'}
\end{eqnarray}
where $\ket{'0'}$ and $\ket{'1'}$ represent eigenstates of the `pointer' observable, $R$, of $\mathbf{M}$. On the standard `eigenvalue-eigenstate rule,' an observable has a definite value if and only if the quantum state is an eigenstate of the observable. The states $ \ket{0}\ket{'0'}$ and $\ket{1}\ket{'1'}$ are eigenstates of $Q$ and $R$ in which these observables have definite values $0, '0'$ or $1, '1'$, respectively, so the value of the observable $Q$ can be inferred from the pointer reading. 

Now, since the unitary quantum dynamics is linear, it follows that if $\mathbf{S}$ is initially in a state $\ket{\psi} = c_{0}\ket{0} + c_{1}\ket{1}$  that is a linear superposition of the eigenstates $\ket{0}$ and $\ket{1}$, then the measurement interaction necessarily yields the evolution:
\begin{equation}
(c_{0}\ket{0} + c_{1}\ket{1})\ket{\mbox{ready}} \rightarrow c_{0}\ket{0}\ket{'0'} + c_{1}\ket{1}\ket{'1'} \label{eqn:meas}
\end{equation}
The entangled state $c_{0}\ket{0}\ket{'0'} + c_{1}\ket{1}\ket{'1'}$ is manifestly not an eigenstate of the observable $Q$, nor is it an eigenstate of the pointer observable $R$.  (Rather, it is an eigenstate of some non-separable observable of the composite system $\mathbf{S} + \mathbf{M}$.) We would like to be able to understand the measurement interaction, in accordance with the probabilistic interpretation of the quantum state, as yielding the event associated with $Q$ taking the value 0 and $R$ taking the value $'0'$, or the event associated with $Q$ taking the value 1 and $R$ taking the value $'1'$, where these distinct pairs of events occur with probabilities $|c_{0}|^{2}$ and  $|c_{1}|^{2}$, respectively. But this is excluded by the linear dynamics and the eigenvalue-eigenstate rule. 

Put this way, the problem involves reconciling the objectivity of a particular measurement outcome with the entangled state at the end of a measurement. 
A solution to the problem would seem to require either modifying the linear dynamics, or modifying the eigenvalue-eigenstate rule, or both. If we insist on the eigenvalue-eigenstate rule, then we must suppose, with von Neumann \cite[p. 351]{Neumann}, that a quantum system $\mathbf{S}$
undergoes a  linear reversible dynamical evolution when $\mathbf{S}$ is not measured, but a quite different nonlinear stochastic irreversible dynamical evolution when an observable, say $Q$, is measured on $S$:
\begin{eqnarray}
c_{0}\ket{0} + c_{1}\ket{1} & \rightarrow & \ket{0},\, \mbox{with probability}\, |c_{0}|^{2} \\
c_{0}\ket{0} + c_{1}\ket{1} & \rightarrow & \ket{1},\, \mbox{with probability}\, |c_{1}|^{2}
\end{eqnarray}
That is, in a measurement of $Q$ on $\mathbf{S}$, the state $\ket{\psi}$ is projected or `collapses' onto one of the eigenstates of $Q$, $\ket{0}$ or $\ket{1}$, with the appropriate probability 
$|c_{0}|^{2}$ or $|c_{1}|^{2}$ given by the Born rule (or, more generally, by Gleason's theorem). This is known as the projection postulate or the `collapse' postulate. The measurement problem then becomes the problem of making sense of this peculiar dual dynamics, in which measured systems behave differently from unmeasured systems. Alternatively, the problem is that of reconciling the unitary dynamical evolution of unmeasured systems with the  non-unitary stochastic dynamical evolution of measured systems. For a measured system  described by a density operator $\rho$ undergoing a unitary transformation $U_{t}$, the evolution  is given by the equation:
\begin{equation}
\rho \rightarrow U^{-1}_{t}\rho U_{t}
\end{equation}
If an observable $A$ with eigenvalues $a_{i}$ is measured on the system, the evolution is given by the equation:
\begin{equation}
\rho \rightarrow \sum_{i}P_{a_{i}}\rho P_{a_{i}} \label{eqn:meas}
\end{equation}
where $P_{a_{i}}$ is the projection operator onto the eigenstate $\ket{a_{i}}$. 

Note that the transition (\ref{eqn:meas}) is just the quantum conditionalization \ref{eqn:luders}, averaged over all possible outcomes of the measurement. The difference between $\rho$ and $\sum_{i}P_{a_{i}}\rho P_{a_{i}}$ is the `irreducible and uncontrollable' measurement disturbance discussed in the previous section.

The GRW theory \cite{GhirardiSEP} solves the measurement problem by introducing a unified stochastic dynamics that covers both sorts of evolution. 
Bohm's theory drops the eigenvalue-eigenstate rule, but the solution to the measurement problem  is also fundamentally dynamical. As we saw in the previous section, a quantum system in Bohm's theory  is represented as a particle with a  trajectory in configuration space, and an observable comes to have a definite value  depending on where the particle moves, under the influence of the guiding field given by the evolving wave function of the system. Dynamical solutions to the measurement problem amend quantum mechanics  in such a way that the loss of
information  in quantum conditionalization  is accounted for dynamically, and the quantum probabilities
are reconstructed dynamically as measurement probabilities. The
quantum probabilities are not regarded as a kinematic feature of
the nonclassical  event structure but are derived dynamically, as artifacts of the measurement
process. 
Even on the Everett interpretation, where Hilbert
space is interpreted as the representation space for a new sort of
ontological entity, represented by the quantum state, and no definite
outcome out of a range of alternative outcomes is selected in a quantum measurement
process (so no explanation is required for such an event), probabilities
arise as a feature of the branching structure that emerges in the dynamical
process of decoherence.

If, instead, we look at the quantum theory as a member of a class of nonclassical `no signaling' theories, in which the state space (considered as a space of multipartite probability distributions) does not have the structure of a simplex, then there is no unique decomposition of mixed states into a convex combination of extremal states, there is no general cloning procedure for an arbitrary extremal state, and there is no measurement in the nondisturbing sense that one has in classical theories, where it is in principle possible, via measurement, to extract enough information about an extremal state to produce a copy of the state without irreversibly changing the state. Hilbert space as a projective geometry
(i.e., the subspace structure of Hilbert space) represents a non-Boolean event space, in which there are built-in, structural probabilistic constraints on correlations between events (associated with the angles between events)---just as in special
relativity the geometry of Minkowski space represents spatio-temporal
constraints on events. These are kinematic, i.e., pre-dynamic,\footnote{See Jannsen \cite{Jannsen2007a} for a similar kinematic-dynamic distinction in the context of special relativity. 'Kinematic' in this sense refers to generic features of systems, independent of the details of the dynamics. In the case of quantum theory, this includes  the association of  Hermitian
operators with observables, the Born probabilities, the von Neumann-L\"{u}ders 
conditionalization rule, and the unitarity constraint on the dynamics, which is related to the event structure via a theorem of Wigner \cite{Wigner1959},\cite{Uhlhorn1963}, but not the interaction dynamics defined by specific Hamiltonians.}  objective probabilistic or information-theoretic constraints
on events to which a quantum dynamics of matter and fields conforms, through its symmetries, just
as the structure of  Minkowski space imposes spatio-temporal kinematic constraints on events to which a relativistic dynamics conforms. In this sense, Hilbert space provides the kinematic framework for the physics of an indeterministic universe, just as Minkowski space provides the kinematic framework for the physics of  a non-Newtonian, relativistic universe. From this perspective, there is no deeper explanation for the quantum phenomena of interference and entanglement than that provided by the structure of Hilbert space, just as there is no deeper explanation for the relativistic phenomena of Lorentz contraction and time dilation than that provided by the structure of Minkowski space.

Pitowsky \cite{PitowskyBetting,Pitowsky07}, has formulated an epistemic Bayesian analysis of quantum probabilities as a logic of partial belief. (See also \cite{SchackBrunCaves2001} and  \cite{CavesFuchsSchack2002}.) By Gleason's theorem, coherence constraints on  `quantum gambles' in the quantum event space defined by the subspace structure of Hilbert space entail  a \emph{unique} assignment of credences, encoded in the quantum state as the credence function of a rational agent. As Pitowsky notes \cite[\S 2.4]{PitowskyBetting}, it would be misleading to characterize this analysis of quantum probabilities as subjective. The correlational constraints of the quantum event space defined by the subspace structure of Hilbert space are objective correlational constraints on events, and the credences encoded in the quantum state are uniquely determined by these probabilistic constraints. Rather, in the sense of Lewis's Principal Principle, Gleason's theorem relates an objective feature of the world, the nonclassical structure of objective chances, to the credence function of a rational agent. As pointed out in the Introduction, objective chances can be understood in the metaphysically `thin' sense as patterns in the Humean mosaic, the totality of all that happens at all times, rather than as irreducible modalities (see \cite{Hoefer2007}, \cite{FriggHoefer2009}).

On this analysis, the quantum state does not have an ontological significance analogous to the ontological significance of an extremal classical state as the `truthmaker' for propositions about the occurrence and non-occurrence of events, i.e., as a  representation of physical reality. Rather, the quantum state is a credence function, a bookkeeping device for keeping track of probabilities.
Conditionalizing on a measurement outcome leads to a
updating of the credence function represented by the quantum state via the von Neumann-L\"{u}ders rule, which---as a non-Boolean or noncommutative version of the classical Bayesian rule for updating an initial probability distribution on new information---expresses the necessary information loss on measurement in a nonclassical theory. Just as Lorentz contraction is a physically real phenomenon
explained relativistically as a kinematic effect of motion in a
non-Newtonian space-time structure, so the change arising in
quantum conditionalization that involves a real loss of information should be understood as a kinematic effect of \emph{any} process
of gaining information of the relevant sort in the non-Boolean probability structure
of Hilbert space, considered as a kinematic framework for an indeterministic physics  (irrespective of the dynamical processes involved in the
measurement process).   If cloning an arbitrary extremal state is impossible, there
can be no deeper explanation for the information loss on conditionalization
than that provided by the structure of Hilbert space as a nonclassical probability theory
or information theory. The definite occurrence of a particular event is constrained
by the kinematic probabilistic correlations represented by the subspace
structure of Hilbert space, and only by these correlations---it is otherwise
free.

In the sense of Shannon's notion of information, which abstracts from semantic features of information and concerns probabilistic correlations between the physical outputs of an information source and a receiver,  this interpretation of the nonclassical features of quantum probabilities is an information-theoretic interpretation.  On this view, what is fundamental in the transition from classical to quantum physics is the recognition that \emph{information in the physical sense has new structural features}, just as the transition from classical to relativistic physics rests on the recognition that space-time is structurally different than we thought. 

From the perspective of the information-theoretic interpretation, there are two distinct measurement problems in quantum mechanics: a  `big' measurement problem and a `small' measurement problem (see \cite{Pitowsky07}, \cite{BubPitowsky2010}). The `big' measurement problem is the problem of explaining how measurements can have definite outcomes, given the unitary dynamics of the theory: it is the problem of explaining \emph{how individual measurement outcomes come about dynamically.} The `small' measurement problem is the problem of accounting for our familiar experience of a classical or Boolean macroworld, given the non-Boolean character of the underlying quantum event space: it is the problem of explaining  the \emph{dynamical emergence of an effectively classical  probability space of macroscopic measurement outcomes} in a quantum measurement process. 

On the information-theoretic interpretation, the `big' measurement problem is a pseudo-problem, a consequence of taking the quantum pure state as the analogue of the classical pure state, i.e., as the `truthmaker' for propositions about the occurrence and non-occurrence of events, rather than as a credence function  associated with the interpretation of Hilbert space as a new kinematic framework for the physics of an indeterministic universe, in the sense that Hilbert space defines  objective probabilistic or information-theoretic constraints on correlations between events.  The `small' measurement problem is a consistency problem that can be resolved by considering the dynamics of the measurement process and the role of decoherence in the emergence of an effectively classical probability space of macro-events to which the Born probabilities refer (alternatively, by considering certain combinatorial features of the probabilistic structure: see Pitowsky \cite[\S4.3]{Pitowsky07}).

In special relativity one has a consistency proof that a dynamical account
of relativistic phenomena in terms of forces is consistent with the kinematic
account in terms of the structure of Minkowski space. An analogous
consistency proof for quantum mechanics would be a dynamical explanation for
the effective emergence of  a classical, i.e., Boolean, event space at the
macrolevel, because it is with respect to the Boolean algebra of the
macroworld that the Born weights of quantum mechanics have empirical cash
value. Here is a sketch of such an explanation (see \cite{BubPitowsky2010}):

Consider the Hilbert space of the entire universe. On the usual view, the quantum analogue of a classical pure state is a quantum pure state represented by a ray or 1-dimensional subspace in Hilbert space. Now, a classical pure state defines a 2-valued homomorphism on the classical Boolean event space. A 2-valued homomorphism---a structure-preserving 0,1 map---partitions events into those that do not occur in the state (mapped onto 0) and those that do occur in the state (mapped onto 1), or equivalently, a 2-valued homomorphism defines a truth-value assignment on the Boolean propositional structure. There is, of course, no 2-valued homomorphism on the quantum event space represented by the non-Boolean algebra of subspaces of Hilbert space,  but a quantum pure state can be taken as distinguishing events that occur at a particular time (events represented by subspaces containing the state, and assigned probability 1 by the state)  from events that don't occur (events represented by subspaces orthogonal to the state, and assigned probability 0 by the state). This leaves all remaining events represented by subspaces that neither contain the state nor are orthogonal to the state (i.e., events assigned a probability $p$ by the state, where $0 < p < 1$) in limbo: neither occurring nor not occurring. The measurement problem then arises as the problem of accounting for the fact that an event that neither occurs nor does not occur when the system is in a given quantum state can somehow occur when the system undergoes a measurement interaction with a macroscopic measurement device---giving measurement a very special status in the theory. 

On the information-theoretic interpretation, the quantum state is a bookkeeping device, a credence function that assigns probabilities to events in alternative Boolean algebras associated with the outcomes of alternative measurement outcomes. The measurement outcomes are macro-events in a \emph{particular} Boolean algebra, and the macro-events that \emph{actually} occur, corresponding to a particular measurement outcome, define a 2-valued homomorphism on \emph{this} Boolean algebra, partitioning all events in the Boolean algebra into those that occur and those that do not occur. \emph{What has to be shown is how this occurrence of events in a particular Boolean algebra is consistent with the quantum dynamics.}

Now, it is a contingent feature of the dynamics of our particular quantum universe that events represented by subspaces of Hilbert space have a tensor product structure that reflects the division of the universe into microsystems (e.g., atomic nuclei), macrosystems (e.g., macroscopic measurement devices constructed from pieces of metal and other hardware), and the environment (e.g., air molecules, electromagnetic radiation). The Hamiltonians characterizing the interactions between microsystems and macrosystems, and the interactions between macrosystems and their environment, are such that a certain relative structural stability emerges at the macrolevel as the  tensor-product structure of events in Hilbert space evolves under the unitary dynamics. Symbolically, an event represented by a 1-dimensional projection operator like $P_{\ket{\psi}} = \ket{\psi}\bra{\psi}$, where
\begin{equation}
\ket{\psi} = \ket{s}\ket{M}\ket{\varepsilon}
\end{equation}
and $s, M, \varepsilon$ represent respectively microsystem, macrosystem, and environment, evolves under the dynamics to $P_{\ket{\psi(t)}}$, where
\begin{equation}
\ket{\psi(t)} = \sum_{k}c_{k}\ket{s_{k}}\ket{M_{k}}\ket{\varepsilon_{k}(t)}, \label{eq:correlation}
\end{equation}
and
\begin{equation}
\ket{\varepsilon_{k}(t)} = \sum_{\nu}\gamma_{\nu}e^{-ig_{k\nu}t}\ket{e_{\nu}}
\end{equation}
if the interaction Hamiltonian $H_{M\varepsilon} $ between a macrosystem and the environment takes the form
\begin{equation}
H_{M\varepsilon} = \sum_{k\gamma}g_{k\nu}\ket{M_{k}}\bra{M_{k}}\otimes\ket{e_{\nu}}\bra{e_{\nu}}
\end{equation}
with the $\ket{M_{k}}$ and the $\ket{e_{k}}$ orthogonal.  That is,  the `pointer' observable $\sum_{k}m_{k}\ket{M_{k}}\bra{M_{k}}$ commutes with $H_{M\varepsilon}$ and so  is a constant of the motion induced by the Hamiltonian $H_{M\varepsilon}$. 

Here $P_{\ket{M_{k}}}$ can be taken as representing, in principle, a configuration of the entire macroworld, and $P_{\ket{s_{k}}}$ a configuration of all the micro-events correlated with macro-events. The dynamics preserves the correlation represented by the superposition $\sum_{k}c_{k}\ket{s_{k}}\ket{M_{k}}\ket{\varepsilon_{k}(t)}$ between micro-events, macro-events, and the environment for the specific macro-events $P_{\ket{M_{k}}}$, even for non\-ortho\-gonal $\ket{s_{k}}$ and $\ket{\varepsilon_{k}}$, but not for non-standard macro-events $P_{\ket{M'_{l}}}$ where the $\ket{M'_{l}}$  are linear superpositions of the $\ket{M_{k}}$. The tri-decomposition $\sum_{k}c_{k}\ket{s_{k}}\ket{M_{k}}\ket{\varepsilon_{k}(t)}$ is unique, unlike the bi\-ortho\-gonal Schmidt decomposition (see Elby and Bub \cite{ElbyBub}). Expressed in terms of  non-standard macrostates $\ket{M'_{l}}$,  the tri-decomposition  takes the form of a linear superposition  in which the non-standard macro-events $P_{\ket{M'_{l}}}$ become correlated with entangled system-environment events represented by  linear superpositions of the form $\sum_{k}c_{k}d_{lk}\ket{s_{k}}\ket{\varepsilon_{k}(t)}$. So for macro-events $P_{\ket{M'_{l}}}$ where the $\ket{M'_{l}}$  are linear superpositions of the $\ket{M_{k}}$, the division into micro-events, macro-events, and the environment  is not preserved.    (See Zurek \cite[p. 052105-14]{Zurek2005}.) 

It is characteristic of the dynamics that correlations represented by (\ref{eq:correlation}) evolve to similar correlations---similar in the sense of preserving the micro-macro-environment division. The macro-events represented by $P_{\ket{M_{k}}}$, at a sufficient level of coarse-graining, can be associated with structures at the macrolevel---the familiar macro-objects of our experience---that remain relatively stable under the dynamical evolution. So a Boolean algebra $\hil{B_{M}}$ of macro-events $P_{\ket{M_{k}}}$ correlated with micro-events $P_{\ket{s_{k}}}$ in (\ref{eq:correlation}) is emergent in the dynamics. Note that the emergent Boolean algebra is not the same Boolean algebra from moment to moment, because the correlation between micro-events and macro-events changes under the dynamical evolution induced by the micro-macro interaction (e.g., corresponding to different measurement interactions).  What remains relatively stable under the dynamical evolution are the \textit{macrosystems} associated with macro-events in correlations of the form (\ref{eq:correlation}), even under a certain vagueness in the  coarse-graining associated with these macro-events: macrosystems like grains of sand, tables and chairs, macroscopic measurement devices, cats, people, galaxies, etc.

It is further characteristic of the dynamics that  the environmental events represented by $P_{\ket{\varepsilon_{k}(t)}}$
very rapidly approach orthogonality, i.e., the `decoherence factor'
\begin{equation}
\zeta_{kk'} = \langle\varepsilon_{k}|\varepsilon_{k'}\rangle = \sum_{\nu}|\gamma_{\nu}|^{2}e^{i(g_{k'\nu}-g_{k\nu})t}
\end{equation}
becomes negligibly small almost instantaneously for $k\neq k'$.  When  the environmental events $P_{\ket{\varepsilon_{k}(t)}}$ correlated with the macro-events $P_{\ket{M_{k}}}$ are effectively orthogonal,  the reduced density operator is effectively diagonal in the `pointer' basis $\ket{M_{k}}$ and there is effectively no interference between elements of the emergent Boolean algebra $\hil{B_{M}}$. It follows that the conditional probabilities of events associated with a subsequent emergent Boolean algebra (a subsequent measurement) are additive on $\hil{B_{M}}$.  (See Zurek \cite[p. 052105-14]{Zurek2005}, \cite{Zurek2003a}.)

The Born probabilities are probabilities of events in the emergent Boolean algebra, i.e.,  the Born probabilities are probabilities of `pointer' positions, the coarse-grained basis selected by the dynamics. Applying quantum mechanics kinematically, say in assigning probabilities to the possible outcomes of a measurement of some observable of a microsystem, we consider the Hilbert space of the relevant degrees of freedom of the microsystem and treat the measuring instrument as simply selecting a Boolean subalgebra of measurement outcomes in the non-Boolean event space of the microsystem on which the Born probabilities are defined as the probabilities of measurement outcomes. In principle, we can include the measuring instrument in a dynamical analysis of the measurement process, where the Born probabilities are derived as the probabilities of the occurrence of events in an emergent Boolean algebra. Since the information loss on conditionalization relative to classical conditionalization is a kinematic feature of the the structure of quantum events, not accounted for by the unitary quantum dynamics, which conforms to the kinematic structure,  such a dynamical analysis does not provide a \emph{dynamical explanation of how individual outcomes come about}. 

This is analogous to the situation in special
relativity, where Lorentz contraction is a kinematic effect of relative
motion that is \emph{consistent} with a dynamical account in terms of
Lorentz invariant forces, but is not explained in Einstein's theory as a dynamical effect (i.e., the dynamics is assumed to have symmetries that
respect Lorentz contraction as a kinematic effect of relative motion). By contrast, in  Lorentz's theory, the contraction is a dynamical effect in a Newtonian
space-time structure, in which this sort of contraction does not arise as a
purely kinematic effect. 
Similarly, in quantum mechanics, the possibility of a dynamical analysis of the measurement process conforming to the kinematic structure of Hilbert space provides a \emph{consistency proof} that the familiar objects of our macroworld behave dynamically in accordance with the kinematic probabilistic constraints on correlations between events. (For an opposing view, see \cite{BrownTimpson2007}, \cite{Duwell2007}.)

Note that the application of decoherence here is to the `small' measurement problem, as the core component of a consistency proof. The usual objection to decoherence as a solution to the measurement problem applies to decoherence as a solution to the `big'
 measurement problem: the objection is that decoherence provides, at best, a FAPP (`for all practical purposes') explanation, to use Bell's pejorative term \cite{BellCH}, of how individual measurement outcomes come about dynamically, in terms of the effective diagonalization of the density matrix, which is no good at all as a solution to the `big' problem. 

\section{Hilbert Space as the Kinematics for an Indeterministic Physics}

The discussion in \S4 outlines an interpretation of Hilbert space as defining the pre-dynamic kinematics of a physical theory of an indeterministic universe: a nonclassical theory of `no signaling' probabilistic correlations, or information in Shannon's sense---just as Minkowski space provides the kinematic framework for the physics of  a non-Newtonian, relativistic universe. No assumption is made about the fundamental `stuff' of the universe. 

So, one might ask, what do macroscopic objects supervene on? In the case of Bohm's theory or the GRW theory, the answer is relatively straightforward: macroscopic objects supervene on particle configurations in the case of Bohm's theory, and on mass density or `flashes' in the case of the GRW theory, depending on whether one adopts the GRWm version or the GRWf version. In the Everett interpretation, macroscopic bodies  supervene on features of the quantum state, which describes an ontological entity. On the information-theoretic interpretation proposed here, macroscopic objects supervene on events defining a 2-valued homomorphism on the emergent Boolean algebra. 

It might be supposed that this involves a contradiction. What is contradictory is to suppose that a correlational event represented by $P_{\ket{\psi(t)}}$ actually occurs, where
$\ket{\psi(t)}$ is a linear superposition $\sum_{k}c_{k}\ket{s_{k}}\ket{M_{k}}\ket{\varepsilon_{k}(t)}$, as well as an event represented by $P_{\ket{s_{k}}\ket{M_{k}}\ket{\varepsilon_{k}(t)}}$ for some specific $k$. On the information-theoretic interpretation, there is a kinematic structure of possible correlations (but no particular atomic correlational event is selected as the `state' in a sense analogous to the pure classical state), and a particular dynamics that preserves certain sorts of correlations, i.e., correlational events of the sort represented by $P_{\ket{\psi(t)}}$ with $\ket{\psi(t)} = \sum_{k}c_{k}\ket{s_{k}}\ket{M_{k}}\ket{\varepsilon_{k}(t)}$ evolve to correlational events of the same form. What can be identified as emergent in this dynamics is an effectively classical probability space: a Boolean algebra with atomic correlational events of the sort represented by orthogonal 1-dimensional subspaces $P_{\ket{s_{k}}\ket{M_{k}}}$, where the probabilities are generated by the reduced density operator obtained by tracing over the environment, when the correlated environmental events are effectively orthogonal.

The quantum dynamics does not describe the (deterministic or stochastic) evolution of the 2-valued homomorphism on which macroscopic objects  supervene to a new 2-valued homomorphism (as in the evolution of a classical state). Rather, the dynamics leads to the relative \emph{stability} of certain event structures at the macrolevel associated with the familiar macrosystems of our experience, and to an emergent effectively classical probability space or Boolean algebra, whose atomic events are correlations between events associated with these macrosystems and micro-events. 

It is consistent with the quantum dynamics to regard the actually occurring events as occurring  with the emergence of the Boolean algebra at the macrolevel. The occurrence of these events is only in conflict with the evolution of the quantum pure state if  the quantum pure state is assumed to have an ontological significance analogous to the ontological significance of the classical pure state as the `truthmaker' for propositions about the occurrence and non-occurrence of events---in particular, if it is assumed  that the quantum pure state partitions all events into events that actually occur, events that do not occur, and events that neither occur nor do not occur, as on the usual interpretation. Here the quantum state, pure or mixed, is understood to represent a credence function: the credence function of a rational agent (an information-gathering entity `in' the emergent Boolean algebra) who is updating probabilities on the basis of events that occur in the emergent Boolean algebra. 

There are other information-theoretic interpretations of quantum mechanics (see \cite{Timpson2008a,Timpsonbook,Timpson2010b} for a critical discussion), the most prominent of which is the information-theoretic interpretation of Fuchs \cite{Fuchs2001b,Fuchs2002a,Fuchs2002b,Fuchs2003}, in which quantum states represent subjective degrees of belief, and the loss of information on measurement is attributed to Bayesian conditionalization as the straighforward refinement of prior degrees of belief in the usual sense, together with a further readjustment of the observer's beliefs, which is required roughly because, as Fuchs puts it \cite[p.8]{Fuchs2002b}: `The world is sensitive to our touch.' For Fuchs, as for de Finetti (see \cite{Galavotti}), physics is an extension of common sense. What does the work in allowing Fuchs' Bayesian analysis of measurement updating  to avoid the measurement problem is, ultimately, an instrumentalist interpretation of quantum probabilities as the probabilities of measurement outcomes. 

By contrast, the information-theoretic interpretation outlined here is proposed as a realist interpretation, in the context of an analysis of nonclassical probabilistic correlations in an indeterministic (non-Boolean) universe, analogous to the analysis of nonclassical spatio-temporal relations in a relativistic universe. A salient feature of this interpretation  is the rejection of one aspect of the measurement problem, the `big' measurement problem, as a pseudo-problem, and the recognition of the `small' measurement problem as a legitimate consistency problem that requires resolution.

\section*{Acknowledgements}

Research supported by  the University of Maryland Institute for Physical Science and Technology. Thanks to Chris Timpson and Claus Beisbart for very helpful critical and editorial comments.

\bibliographystyle{plain}
\bibliography{qprobs}

\end{document}